\documentclass[onecolumn,aps,pra]{revtex4}
\usepackage{bm}         
\usepackage{ifpdf}
\usepackage{amssymb,lineno,amsfonts}
\usepackage{graphicx}   
\usepackage{bbm}
\usepackage{mathrsfs}
\usepackage{upgreek}
\usepackage{mathtools}
\usepackage{epstopdf}
\usepackage{setspace}
\usepackage{hyperref}
\usepackage[matrix,frame,arrow]{xypic}
\usepackage{rotating}
\usepackage{float}
\usepackage{array}
\usepackage{color} 
\definecolor{med-blue}{RGB}{25,25,112} 
\hypersetup{colorlinks, linkcolor={red},citecolor={blue}, urlcolor={blue}}

\newcommand{\ket}[1]{\vert{#1}\rangle}
\newcommand{\bra}[1]{\langle{#1}\vert}

\newcommand{\inpr}[2]{\langle{#1}\vert{#2}\rangle}

\newcommand{\abs}[1]{\vert{#1}\vert}

\begin{document}

\vspace*{1cm}
\title{Design and experimental realization of an optimal scheme for  teleportion of an $n$-qubit quantum state}
\author{Mitali Sisodia$^1$, Abhishek Shukla$^1$, Kishore Thapliyal$^1$ and Anirban Pathak$^1$} 
\email{anirban.pathak@gmail.com} 
\affiliation{$^1$Jaypee Institute of Information  Technology, A-10, Sector 62, Noida, UP 201307, India} 

\begin{abstract}
An explicit scheme (quantum circuit) is designed for the teleportation of an $n$-qubit  quantum state. It is established that the proposed scheme requires an optimal amount of quantum resources, whereas larger amount  of quantum resources has been used in a large number of recently reported teleportation schemes for the quantum states which can be viewed as special cases of the general $n$-qubit state considered here. A trade off between our knowledge about the quantum state to be teleported and the amount of quantum resources required for the same is observed. A proof of principle experimental realization of the proposed scheme (for a 2-qubit state) is also performed using 5-qubit superconductivity-based IBM quantum computer. Experimental results show that the state has been teleported with high fidelity. Relevance of the proposed teleportation scheme has also been discussed in the context of controlled, bidirectional, and bidirectional-controlled state teleportation. 

\end{abstract}

\maketitle

\section{Introduction} \label{intro}
Perfect teleportation of an arbitrary quantum state using a classical channel would require an infinite amount of classical resources. However,  it is possible to teleport an arbitrary quantum state with unit fidelity using a quantum channel and a few bits of classical communication. As perfect teleportation does not have a classical analogue, schemes for quantum teleportation (QT) drew considerable attention of the quantum communication community since its introduction in 1993 by Bennett et al.,  \cite{bennett1993teleporting}. As a consequence, a large number of modified QT schemes have been proposed. In the present scenario, these schemes span a wide spectrum of quantum communication science. To elucidate this point, we may mention a few schemes of quantum communication tasks,  which  may be viewed as modified schemes for  QT.  This set of quantum communication tasks includes--remote state preparation (RSP) \cite{pati2000minimum}, controlled teleportation (CT) \cite{karlsson1998quantum,pathak2011efficient}, bidirectional state teleportation (BST) \cite{huelga2001quantum}, bidirectional controlled state  teleportation (BCST) \cite{zha2013bidirectional,shukla2013bidirectional,thapliyal2015general,thapliyal2015applications}, quantum information splitting (QIS) \cite{hillery1999quantum,nie2011quantum}, quantum secret sharing (QSS) \cite{hillery1999quantum}, hierarchical versions of these schemes \cite{shukla2013hierarchical,mishra2015integrated,shukla2016hierarchical}, and many more. Many of these schemes have very interesting applications (for details see \cite{P13}). For example, we may mention that every scheme of  BST can be used to design quantum remote control  \cite{huelga2001quantum}.  It is also worth noting here that although teleportation (and most of its variants) in its original form is not a secure communication scheme, it can be used as a primitive for secure quantum communication. On top of that, there are proposals to employ teleportation in quantum repeaters to enhance the feasibility of quantum communication, and in ensuring security against an eavesdropper's attempt to encroach the private spaces of legitimate users devising trojan-horse attack  \cite{lo1999unconditional}. This wide applicability of QT and its variants and the fact that quantum resources are costly (for example, preparation and maintenance of an $n$-qubit entangled state is extremely difficult for large $n$) have motivated us to investigate whether the recently proposed schemes \cite{li2016quantum,hassanpour2016bidirectional,da2007teleportation,li2016asymmetric,song2008controlled,cao2005teleportation,muralidharan2008quantum,tsai2010teleportation,nie2011quantum,tan2016deterministic,wei2016comment,li2016quantum1,yu2013teleportation,nandi2014quantum} are using an  optimal amount of quantum resources? If not, how to reduce the amount of quantum resources to be used? An effort to answer these questions has led to the present work, where we aim to propose a scheme for teleportation of an $n$-qubit quantum state using an optimal amount of quantum resources  and to experimentally realize a particular case of the proposed scheme using 5-qubit IBM quantum computer. Before we proceed to describe the findings of the present work, we would like to elaborate a bit on what makes it fascinating to work on teleportation even after  almost quarter century of its introduction.

As a natural generalization of QT schemes, protocols for teleportion of multi-qubit states have also been proposed. For example, Chen et al. \cite{chen2006general} have presented a general form of a genuine multipartite entangled channel. Their work encapsulates the essence that the bipartite nature of the channel is sufficient for teleportation of multi-qubit quantum states. Furthermore, they  concluded that QT of an arbitrary $n$-qubit state can be achieved by performing $n$ rounds of Bennett et al.'s protocol \cite{bennett1993teleporting} for QT (one for each qubit).  Later, this scheme was extended to CT of an arbitrary $n$-qubit quantum state \cite{ man2007genuine}. Along the same line, a BST and a BCST schemes may be designed for teleporting two arbitrary multi-qubit states, one each by Alice and Bob. Specifically, a quantum state suitable as a quantum channel for a CT (BCST) scheme should essentially reduce to a useful quantum channel for QT (BST) after the controller's measurement (see \cite{thapliyal2015general} for detail discussion).  Most of these results related to teleporation of multi-qubit states are theoretical in nature. In these  schemes,  multi-qubit entangled states are used  without giving much attention to the possibility of  experimental generation and maintenance of such entangled states. On the other hand,  first experimental realization of Bennett et al.'s protocol was reported by Bouwmeester et al., \cite{bouwmeester1997experimental} in 1997 using photons. Later on, a number of realizations using other architectures have also been reported \cite{bouwmeester1997experimental,nielsen1998complete,furusawa1998unconditional,zhao2004experimental,riebe2004deterministic,barrett2004deterministic}.   
However, hardly any proposal for teleportation of multi-qubit quantum states have been tested experimentally  becuase experimental realization of those schemes would require quantum resources that are costly. This observation has further motivated us to design a general teleportation scheme that would require  lesser amount of quantum resource and/or such resources that can be generated and maintained easily using available technology.

It may be noted that an optimized scheme for multipartite QT must involve optimization of both procedure and resources. Optimization of the procedure demands use of those states as quantum channel, which can be prepared easily and are least affected by decoherence; while the optimization of resources that the scheme should exploit/utilize all the channel qubits that are available for performing QT. The results of Ref. \cite{chen2006general} can be viewed as an optimization of procedure (as Bell states can be easily prepared and are less prone to decoherence in comparison with the multipartite entangled states). The importance of such strategies becomes evident in experimental scenarios, say  when we wish to realize QT in a quantum network designed for quantum internet \cite{sun2016quantum}. Naturally, an  optimized QT scheme would improve the performance of such a quantum internet.  Another kind of optimization has been attempted in some recent works. Specifically, efforts have been made to form teleportation channel (having lesser number of entangled qubits) suitable for the teleportation of specific types of unknown quantum states \cite{li2016quantum,hassanpour2016bidirectional,da2007teleportation,li2016asymmetric,song2008controlled,cao2005teleportation,muralidharan2008quantum,tsai2010teleportation,nie2011quantum,tan2016deterministic,wei2016comment,li2016quantum1,yu2013teleportation,nandi2014quantum}.  For example, in \cite{li2016quantum}, the 3-qubit and the 4-qubit unknown quantum states have been teleported using 4 and 5-qubit cluster states, respectively.  An extended list of these complex states and the corresponding channels are given in Table \ref{tab:lit}.  In fact, some of this set of schemes has exploited the fact that some of the probability amplitudes in the state to be teleported are zero and  a QT scheme essentially transfers the unknown probability amplitudes to distant qubits. We noticed that the quantum resources used in these protocols are not optimal and most of the cases involve redundant qubits.  Keeping these in mind, here, we set ourselves the task to minimize the number of these qubits exploiting the available information regarding the mathematical structure of the quantum state to be teleported. Specifically, in what follows, we would propose an efficient and optimal (uses minimum number of Bell states as quantum channel) scheme to teleport a multi-qubit state of specific form.  Further, it would be established that the proposed scheme is simple in nature and can be extended to corresponding CT and BCST schemes.

Actual relevance of an optimized scheme lies in the experimental realization of the scheme only. An interesting window for experimental realization of the schemes of quantum computation and communication in general and optimized schemes in particular has been opened up recently, when IBM provided free access to a 5-qubit superconducting quantum computer by placing it in cloud \cite{IBMQE,devitt2016performing}. This has provided a platform for experimental testing of various proposals for quantum communication and computation. In the present work, we have used IBM quantum computer to experimentally realize the optimal scheme designed here. Specifically, we have successfully implemented the optimal quantum circuit designed for teleportation of a 2-qubit state. The experiment performed is a proof of principle experiment as the receiver and the sender are not located at two distant places, but it shows successful teleportation with very high fidelity and paves away the path for future realizations of the proposed scheme using optical elements.

Rest of the paper is organized as follows.  In Section \ref{mun}, we propose a scheme for the teleporation of a multi-qubit quantum state having $m$ unknown coefficients using optimal quantum resources. We also discuss a specific case of this scheme which corresponds to QT of a two-unknown multi-qubit quantum state using a Bell state as quantum channel. In Section \ref{cbt}, we  describe optimal schemes for controlled unidirectional and bidirectional state teleportation the quantum states. 
Further, in Section \ref{ex}, an experimental realization of the proposed scheme is reported using the 5-qubit IBM quantum computer available on cloud. Finally, we conclude the paper in section \ref{con}.

\section{Teleportation of an $n$-qubit state with $m$ unknown coefficients} \label{mun}

Consider an $n$-qubit quantum state to be teleported as

\begin{equation}
|\psi\rangle=\sum_{i=1}^{m}\alpha_{i}\ket{x_{i}},  \label{eq:psi}
\end{equation}
where $\alpha_i$s are the probability amplitudes ensuring normalization $\sum_{i=1}^{m}\vert\alpha_{i}\vert^{2}=1$. Additionally, $x_i$s are mutually orthogonal to each other, i.e., $\langle x_i|x_j\rangle=\delta_{ij}\,\forall\,\left(1<i,j<m\right)$. Therefore, one may note that $x_i$s are the elements of an $n$-qubit orthogonal basis iff $m\leq2^n$. For example, we may consider 3-qubit quantum states   $\ket{\xi_1}=\alpha_1|000\rangle+\alpha_2|111\rangle$ and $\ket{\xi_2}=\alpha_1|000\rangle+\alpha_2|011\rangle+\alpha_3|100\rangle+\alpha_4|111\rangle$, teleportation schemes for which were proposed in Refs. 
in \cite{yu2013teleportation} and \cite{wei2016comment}, respectively. For both the states  $n=3$, but we can easily observe that $m=2$ for $\ket{\xi_1}$ and $m=4$ for $\ket{\xi_2}$. In what follows, we will establish that because of this difference in the values of $m$, teleportation of $\ket{\xi_1}$ would require only one Bell state, whereas that of $\ket{\xi_2}$ would require 2 Bell states.

\begin{figure}
	\begin{center}			\includegraphics[scale=0.52]{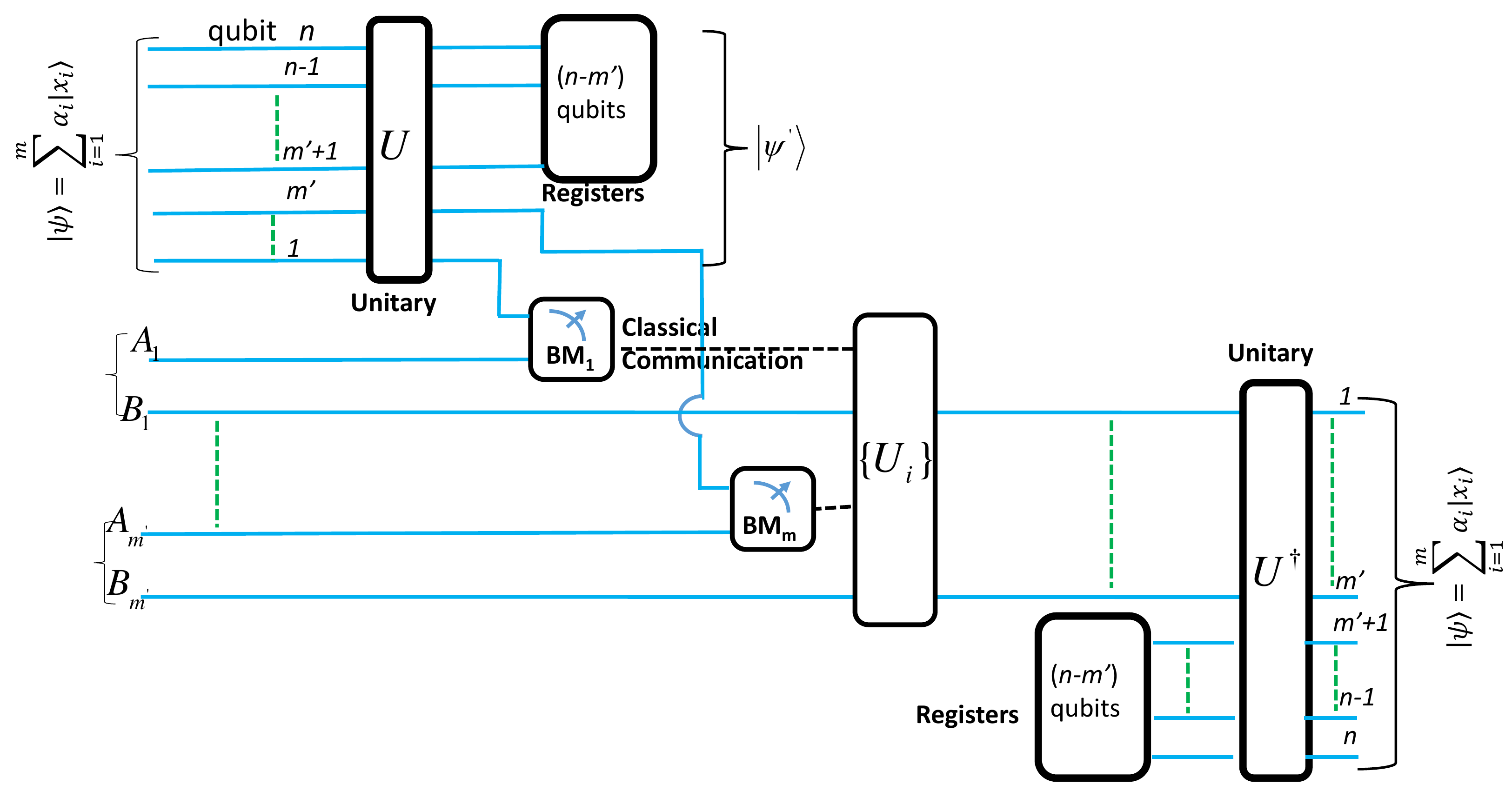}	\end{center}\protect\caption{\label{fig:mun} (Color online) The circuit for teleportation of an $n$-qubit quantum states having  $m$ unknown coefficients. Here, $m^\prime=\left\lceil \log_2{m}\right\rceil $ is the number of Bell states used which depends on the number of unknowns $m$. Also, $U_i$ is the unitary operation Bob has to apply to reconstruct the teleported quantum state.}
		\end{figure}

An arbitrary $n$-qubit quantum state possesses $2^n$ superposition, i.e., $m=2^n.$ Here, we set ourselves a task to teleport state $\ket{\psi}:\, m<2^n$ using optimal quantum resources (i.e., using minimum number of qubits in the multiqubit entangled state used as quantum channel). To do so, we will transform the state to be teleported (say, $|\psi\rangle$) to a quantum state of preferred form (say, $|\psi^{\prime}\rangle$). Specifically, the central idea of our scheme lies in finding out a unitary $U$, which transforms state $\ket{\psi}$ into $\ket{\psi^\prime}$, i.e., $U\ket{\psi}= \ket{\psi^\prime}$, such that

\begin{equation}
|\psi^\prime\rangle=\sum_{i=1}^{m}{\alpha_{i}^{\prime}|y_{i}\rangle}. \label{eq:psiD}
\end{equation}
Here, $|\psi^\prime\rangle $ is a unique $n$-qubit quantum state which is written using only $m$ out of total $n$ elements of the computational basis $\{y_i\}$. Specifically, the unitary is expected to possess a one-to-one map for each element of $\{x_i\} \rightarrow \{y_i\}$. 
As shown in Figure \ref{fig:mun}, here we prefer $\ket{y_i} = \ket{0}^{n-m^{\prime}} \otimes \ket{\widetilde{m}_i}$, where $m^{\prime}=\left\lceil \log_2{m}\right\rceil $ and $\widetilde{m}_i$ is the binary equivalent of decimal value $m_i$ in an $m^{\prime}$-bit binary string. This step transforms $m$ elements of $\ket{\psi}$ with non-zero projections in Eq. \ref{eq:psi} to that of $m^\prime$ elements of  $\ket{\psi^\prime}$ in Eq. \ref{eq:psiD}. 

However, to design the unitary able to accomplish such a task the map $\{x_i\} \rightarrow \{y_i\}$ should exist between all the elements (both possessing either zero or non-zero projection in $\ket{\psi}$ and $\ket{\psi^\prime})$ in both the basis. In other words, the unitary $U$ mapping the basis elements from $\{x_i\}$ to $\{y_i\}$ required for the desired transformation would be

\begin{equation}
U=\sum_{i=1}^{2^n}|y_{i}\rangle\langle x_{i}|
\end{equation}
Being a $2^n$ dimensional computational basis $\left\{y_{i}\right\}$ is already known
while state $\ket{\psi}$ reveals only $m$ orthogonal vectors of basis $\{x_i\}$. Therefore, the remaining $\left(2^n-m\right)$ elements of basis $\{x_i\}$ can be obtained by Gram-Schmidt procedure, such that the elements of $\{x_i\}$ follow the completeness relation, i.e., $\sum_{i=1}^{2^n}{\sum}|x_{i}\rangle\langle x_{i}|=\mathbb{I}$. 

The obtained quantum state $\ket{\psi^\prime}$ possesses the first ${n-m^{\prime}}$ qubits in $\ket{0}$, while the remaining $m^{\prime}$ qubits hold the complete information of $\ket{\psi}$. Therefore, our task reduces to a teleportation of an $m^{\prime}$-qubit quantum states using an optimal amount of quantum resources. An $m^{\prime}$-qubit quantum state can be teleported either by using at least $2m^{\prime}$-qubit entangled state or $m^{\prime}$ Bell states, one for teleporting each qubit \cite{chen2006general}. 

Preparing a multi-qubit entangled state is relatively expensive and  such a state is more prone to decoherence than a two qubit entangled state. For the reason, we prefer the second strategy and select Bell states as a quantum channel (see  Figure \ref{fig:mun}). Once the quantum state $\ket{\psi^\prime}$ is reconstructed at Bob's port, he would require to perform a unitary operation $U^{\dagger}$. 

For the sake of completeness of the paper we may summarize teleportation of an arbitrary $m^{\prime}$-qubit quantum state in the following steps:

\begin{enumerate}
	\item The $m^{\prime}$-qubit unknown state to be teleported (whose qubits are indexed by $1,2,\cdots, m^{\prime}$) and the first qubits of each $m^{\prime}$ Bell states (indexed by $A_{1}, A_{2}, \cdots, A_{m^{\prime}}$) are with Alice and all the second qubits (indexed by $B_{1}, B_{2}, \cdots, B_{m^{\prime}}$) are with Bob.	
	\item Alice performs pairwise Bell measurements on her qubits ($i, A_{i}$) and finally announces $m^{\prime}$ measurement outcomes.
	
	\item Bob applies Pauli operations on each qubit in his possession depending upon the measurement outcome of Alice (see Table 1 of Ref. \cite{thapliyal2015applications} for detail). At the end of this step Bob obtains the $m^{\prime}$-qubit unknown quantum state that Alice has teleported.
\end{enumerate}

Once Bob has access to the $m^{\prime}$-qubit unknown state  and knowledge of the unitary $U$ Alice has applied, he prepares $n-m^{\prime}$ qubits in $\ket{0}$. Finally, he applies $U^{\dagger}$ to reconstruct the unknown
	quantum state $\begin{array}{lcl}
	U^{\dagger}|\psi^{\prime}\rangle & = & |\psi\rangle\end{array}$

In Table \ref{tab:lit}, we give unitary operations involved in teleportation of various multiqubit states with different number of unknowns using our scheme. Teleportation of these states using relatively fragile and expensive quantum resources have been reported in the recent past. 
To be specific, our technique can be used to teleport any  quantum state having two unknown coefficients \cite{da2007teleportation,cao2005teleportation,tsai2010teleportation,yu2013teleportation,li2016quantum}  using a single Bell state, irrespective of the number of physical qubits present in the state. In contrast, in the  existing literature (cf. Columns 2 and 3 of Table \ref{tab:lit}) it is observed that the number of qubits used in the quantum channel  increases with  the increase in number of qubits to be teleported. Similarly, a quantum state having four-unknown coefficients can be teleported only using two Bell states, unlike the higher dimensional entangled states used in \cite{wei2016comment,li2016quantum1,tan2016deterministic}. This clearly establishes that our scheme allows one to circumvent the use of redundant qubits and complex entangled states that are used until now, and thus the present proposal increases the possibility of experimental realization.

\begin{center}
\begin{table}
\begin{tabular}{|>{\centering}p{2.6cm}|>{\centering}p{1.6cm}|>{\centering}p{2.9cm}|>{\centering}p{1.8cm}|>{\centering}p{7.4cm}|}
	\hline 
	 Quantum state to be teleported   & Number of qubits in the state to be teleported  & Quantum state used as quantum channel and corresponding reference  & Number of Bell states required in our scheme to teleport the state & Unitary to be applied by Alice \tabularnewline
	\hline
		$\alpha|a_{0}\rangle+\beta|a_{3}\rangle$ & 2-qubit  &  4-qubit CS \cite{da2007teleportation},  3-qubit W class state  \cite{cao2005teleportation} & 1 & $|a_{0}\rangle\langle a_{0}|+|a_{1}\rangle\langle a_{3}|+|a_{2}\rangle\langle a_{1}|+|a_{3}\rangle\langle a_{2}|$\tabularnewline
	\hline 
		$\alpha|a_{1}\rangle+\beta|a_{2}\rangle$ & 2-qubit  & 3-qubit GHZ like state \cite{tsai2010teleportation}   & 1  & $|a_{0}\rangle\langle a_{1}|+|a_{1}\rangle\langle a_{2}|+|a_{2}\rangle\langle a_{0}|+|a_{3}\rangle\langle a_{3}|$\tabularnewline
	\hline
		$\alpha|a_{0}\rangle+\beta|a_{7}\rangle$ & 3-qubit & 4-qubit CS \cite{yu2013teleportation} & 1 & $\begin{array}{l}
	|a_{0}\rangle\langle a_{0}|+|a_{1}\rangle\langle a_{7}|
	+|a_{2}\rangle\langle a_{1}|+|a_{3}\rangle\langle a_{2}|\\
	+|a_{4}\rangle\langle a_{3}|+|a_{5}\rangle\langle a_{4}|
	+|a_{6}\rangle\langle a_{5}|+|a_{7}\rangle\langle a_{6}|
	\end{array}$\tabularnewline
	\hline 
	$\begin{array}{l}
	\alpha(|a_{0}\rangle+|a_{3}\rangle)\\
	+\beta(|a_{4}\rangle+|a_{7}\rangle)
	\end{array}$ & 3-qubit  & 4-qubit CS \cite{li2016quantum} &  1	 & $\begin{array}{l}
	\frac{1}{\sqrt{2}}(|a_{0}\rangle\langle(|a_{0}+a_{3})|+|a_{1}\rangle\langle a_{4}+a_{7})|\\
	+|a_{2}\rangle\langle(a_{1}+a_{2})|+|a_{3}\rangle\langle(a_{5}+a_{6})|\\
	+|a_{4}\rangle\langle(a_{0}-a_{3})|+|a_{5}\rangle\langle(a_{4}-a_{7})|\\
	+|a_{6}\rangle\langle(a_{1}-a_{2})|+|a_{7}\rangle\langle(a_{5}-a_{6})|)
	\end{array}$\tabularnewline
	\hline 
	  
	$\begin{array}{l}
	\alpha(|a_{0}\rangle+|a_{7}\rangle)\\
	+\beta(|a_{13}\rangle+|a_{10}\rangle)
	\end{array}$  & 4-qubit  & 5-qubit CS \cite{li2016quantum} & 1 & $\begin{array}{l}
	\frac{1}{\sqrt{2}}(|a_{0}\rangle\langle|a_{0}+a_{7}|+|a_{1}\rangle\langle(a_{13}+a_{10})|\\
	+|a_{2}\rangle\langle(a_{1}+a_{2})|+|a_{3}\rangle\langle(a_{3}+a_{4})|\\
	+|a_{4}\rangle\langle(a_{5}+a_{6})|+|a_{5}\rangle\langle(a_{8}+a_{9})|\\
	+|a_{6}\rangle\langle(a_{11}+a_{12}|+|a_{7}\rangle\langle(a_{14}+a_{15})|\\
	+|a_{8}\rangle\langle(a_{0}-a_{7})|+|a_{9}\rangle\langle(a_{13}-a_{10})|\\
	+|a_{10}\rangle\langle(a_{1}-a_{2})|+|a_{11}\rangle\langle(a_{3}-a_{4})|\\
	+|a_{12}\rangle\langle(a_{5}-a_{6})|+|a_{13}\rangle\langle(a_{8}-a_{9})|\\
	+|a_{14}\rangle\langle(a_{11}-a_{12})|+|a_{15}\rangle\langle(a_{14}-a_{15})|)
	\end{array}$\tabularnewline
	\hline 
	
	$\begin{array}{l}
	\alpha|a_{0}\rangle+\beta|a_{3}\rangle\\
	+\gamma|a_{4}\rangle+\delta|a_{7}\rangle
	\end{array}$  & 3-qubit & 5-qubit state \cite{wei2016comment} &  2  & $\begin{array}{l}
	|a_{0}\rangle\langle a_{0}|+|a_{1}\rangle\langle a_{3}|
	+|a_{2}\rangle\langle a_{4}|+|a_{3}\rangle\langle a_{7}|\\
	+|a_{4}\rangle\langle a_{1}|+|a_{5}\rangle\langle a_{2}|
	+|a_{6}\rangle\langle a_{5}|+|a_{7}\rangle\langle a_{6}|
	\end{array}$\tabularnewline
	\hline 
	
		$\begin{array}{l}
	\alpha|a_{0}\rangle+\beta|a_{3}\rangle\\
	+\gamma|a_{12}\rangle+\delta|a_{15}\rangle
	\end{array}$ & 4-qubit  & 6-qubit CS \cite{li2016quantum1} & 2
	& $\begin{array}{l}
	|a_{0}\rangle\langle a_{0}|+|a_{1}\rangle\langle a_{3}|
	+|a_{2}\rangle\langle a_{12}|+|a_{3}\rangle\langle a_{15}|\\
	+|a_{4}\rangle\langle a_{1}|+|a_{5}\rangle\langle a_{2}|
	+|a_{6}\rangle\langle a_{4}|+|a_{7}\rangle\langle a_{5}|\\
	+|a_{8}\rangle\langle a_{6}|+|a_{9}\rangle\langle a_{7}|
	+|a_{10}\rangle\langle a_{8}|+|a_{11}\rangle\langle a_{9}|\\
	+|a_{12}\rangle\langle a_{10}|+|a_{13}\rangle\langle a_{11}|
	+|a_{14}\rangle\langle a_{13}|+|a_{15}\rangle\langle a_{14}|
	\end{array}$\tabularnewline
	\hline

	 $\begin{array}{l}
	\alpha|a_{0}\rangle+\beta|a_{15}\rangle\\
	+\gamma|a_{63}\rangle+\delta|a_{48}\rangle
	\end{array}$ & 6-qubit  & 6-qubit CS \cite{tan2016deterministic} & 2 & $\begin{array}{l} |a_{0}\rangle\langle a_{0}|+|a_{1}\rangle\langle a_{15}|+|a_{2}\rangle\langle a_{63}|+|a_{3}\rangle\langle a_{48}|\\
	+|a_{4}\rangle\langle a_{1}|+|a_{5}\rangle\langle a_{2}|+|a_{6}\rangle\langle a_{3}|+|a_{7}\rangle\langle a_{4}|\\
	+|a_{8}\rangle\langle a_{5}|+|a_{9}\rangle\langle a_{6}|+|a_{10}\rangle\langle a_{7}|+|a_{11}\rangle\langle a_{8}|\\ 
	+|a_{12}\rangle\langle a_{9}|+|a_{13}\rangle\langle a_{10}|+|a_{14}\rangle\langle a_{11}|+|a_{15}\rangle\langle a_{12}|\\
	+|a_{16}\rangle\langle a_{13}|+|a_{17}\rangle\langle a_{14}|+|a_{18}\rangle\langle a_{16}|+|a_{19}\rangle\langle a_{17}|\\
	+|a_{20}\rangle\langle a_{18}|+|a_{21}\rangle\langle a_{19}|+|a_{22}\rangle\langle a_{20}|+|a_{23}\rangle\langle a_{21}|\\
	+|a_{24}\rangle\langle a_{22}|+|a_{25}\rangle\langle a_{23}|+|a_{26}\rangle\langle a_{24}|+|a_{27}\rangle\langle a_{25}|\\
	+|a_{28}\rangle\langle a_{26}|+|a_{29}\rangle\langle a_{27}|+|a_{30}\rangle\langle a_{28}|+|a_{31}\rangle\langle a_{29}|\\
	+|a_{32}\rangle\langle a_{30}|+|a_{33}\rangle\langle a_{31}|+|a_{34}\rangle\langle a_{32}|+|a_{35}\rangle\langle a_{33}|\\
	+|a_{36}\rangle\langle a_{34}|+|a_{37}\rangle\langle a_{35}|+|a_{38}\rangle\langle a_{36}|+|a_{39}\rangle\langle a_{37}|\\
	+|a_{40}\rangle\langle a_{38}|+|a_{41}\rangle\langle a_{39}|+|a_{42}\rangle\langle a_{40}|+|a_{43}\rangle\langle a_{41}|\\
	+|a_{44}\rangle\langle a_{42}|+|a_{45}\rangle\langle a_{43}|+|a_{46}\rangle\langle a_{44}|+|a_{47}\rangle\langle a_{45}|\\
	+|a_{48}\rangle\langle a_{46}|+|a_{49}\rangle\langle a_{47}|+|a_{50}\rangle\langle a_{49}|+|a_{51}\rangle\langle a_{50}|\\
	+|a_{52}\rangle\langle a_{51}|+|a_{53}\rangle\langle a_{52}|+|a_{54}\rangle\langle a_{53}|+|a_{55}\rangle\langle a_{54}|\\
	+|a_{56}\rangle\langle a_{55}|+|a_{57}\rangle\langle a_{56}|+|a_{58}\rangle\langle a_{57}|+|a_{59}\rangle\langle a_{58}|\\
	+|a_{60}\rangle\langle a_{59}|+|a_{61}\rangle\langle a_{60}|+|a_{62}\rangle\langle a_{61}|+|a_{63}\rangle\langle a_{62}|
	\end{array}$\tabularnewline
	\hline

	\end{tabular} 
\caption{\label{tab:lit} A list of quantum states  and the resources used to teleport them in the recent past. The unitary required to decrease the number of entangled qubits to be used as quantum channel is also mentioned. Here, CS stands for cluster state, and $a_{i}$ corresponds to the binary value of decimal number $i$ expanded upto $k$ digits if the entry in the second column of the same row is $k$ qubit. For example, entry in first row second column is 2 qubit, so the state to be teleported (see first row first column) would be $\alpha|a_{0}\rangle+\beta|a_{3}\rangle=\alpha|00\rangle+\beta|11\rangle$. Thus, $a_{0}$ is expanded here up to two digits, but in the third row $a_{0}$ will be expanded as $|000\rangle$, i.e., up to three digits as the entry in third row second column is 3 qubit. Naturally, in this particular case, state to be teleported is $\alpha|a_{0}\rangle+\beta|a_{7}\rangle=\alpha|000\rangle+\beta|111\rangle$. In our scheme, both $\alpha|00\rangle+\beta|11\rangle$ and $\alpha|000\rangle+\beta|111\rangle$ can be teleported using single Bell state, as the resource required depends only on the number of unknown coefficients.}
\end{table}
\end{center}
\hspace{0.5cm}

\subsection{Teleportation of state of type $\ket{\psi} = \alpha \ket{x_i} + \beta \ket{x_j}$} \label{2un}
 
As an explicit example of the proposed scheme, consider an $n$-qubit state with only two unknowns, i.e., 
\begin{equation}
\ket{\psi} = \alpha \ket{x_i} + \beta \ket{x_j},
\end{equation}
such that, $\inpr{x_i}{x_j} = \delta_{ij}$ and ${\abs{\alpha}}^{2} + {\abs{\beta}}^{2}= 1$. Here, $x_i$ and $x_j$ are the elements of some $2^{n}$ dimensional basis set. Our task is to teleport state $\ket{\psi}$ using optimal quantum resources (i.e., using minimum number of entangled qubits in quantum channel). As mentioned previously the minimum number of Bell states required in the quantum channel for this kind of state would be $ \left\lceil \log_2{2}\right\rceil =1$. 

Therefore, we will transform state $\ket{\psi}$ into $\ket{\psi^\prime}$, such that $U\ket{\psi}= \ket{\psi^\prime}$, such that
\begin{equation}
\ket{\psi^\prime} = \alpha^\prime \ket{y_i} + \beta^\prime  \ket{y_j} 
\end{equation}
with, $\inpr{y_i}{y_j}= \delta_{ij}$ as $y_{i}$ and $y_{j}$ are the the elements of computational basis. For the simplest choice of $\ket{\psi^\prime}$, we choose $y_{i} =  {\ket{0}}^{n-1} \otimes \ket{0}$ and $y_{j} =   {\ket{0}}^{n-1}\otimes\ket{1}$. 

\begin{figure}
	\begin{center}		\includegraphics[scale=0.52]{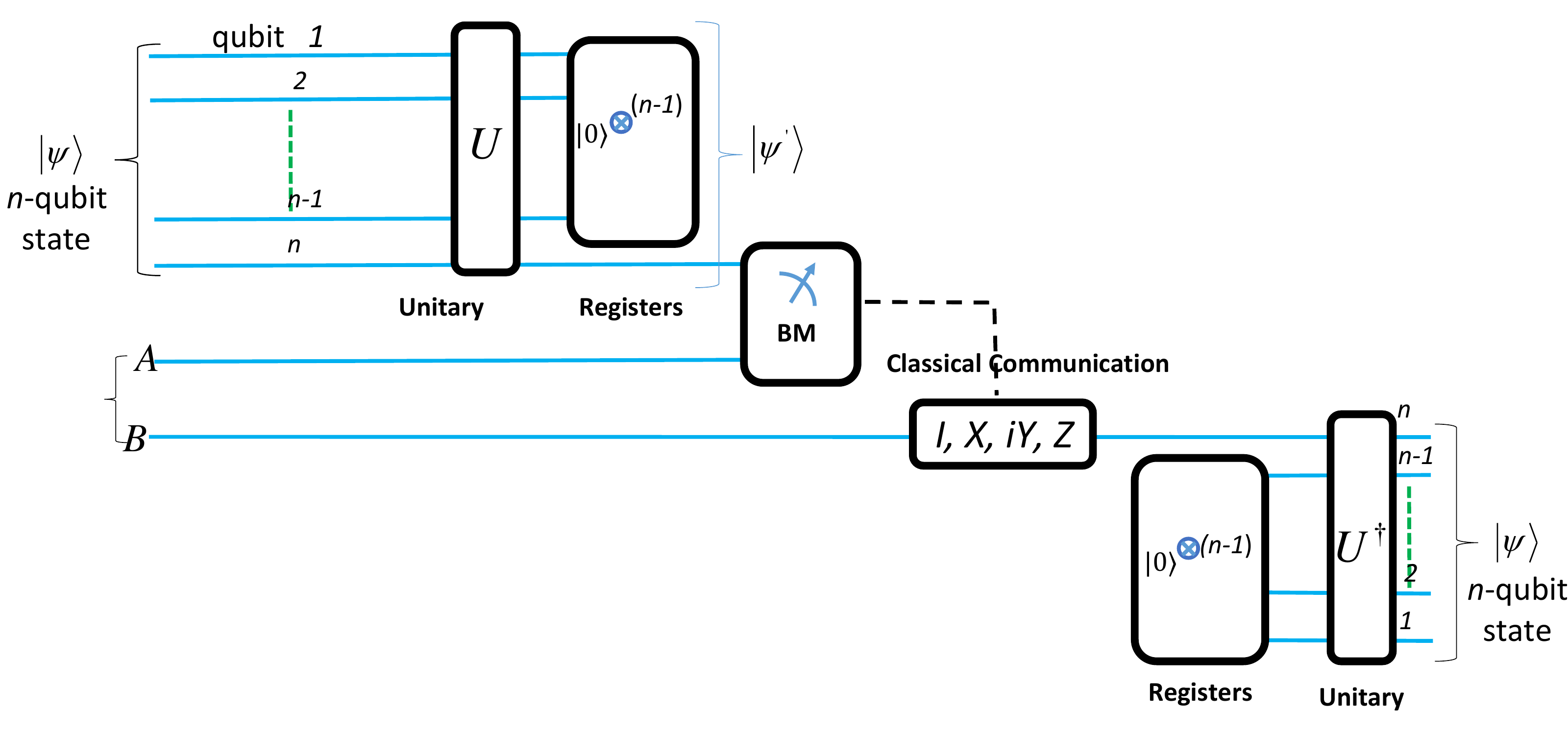}\end{center}\protect\caption{\label{fig:2un}(Color online) As an explicit example, a quantum circuit is shown that performs teleportation of an $n$-qubit quantum state having 2 unknown coefficients using only a single Bell state.}
		\end{figure}

Now, we will show that it is possible to teleport state $\ket{\psi^\prime}$ using one $e$ bit (Bell state) and classical resource.
The quantum circuit for teleporting state $\ket{\psi}$ is given in Figure \ref{fig:2un}. The first part of the quantum circuit shows  transformation of the state $\ket{\psi}$ to the state $\ket{\psi^\prime}$ while the second part of the circuit contains standard scheme for teleporting an arbitrary single qubit state. The third part of the circuit involves application of the unitary ${\mathrm U}^{\dagger}$ to transform the reconstructed state $\ket{\psi^\prime}$ into the unknown state $\ket{\psi}$ to be teleported.

\section{Controlled and Bidirectional teleportation with optimal resource}\label{cbt}
Controlled teleportation of a single qubit involves a third party (Charlie)  as supervisor and hence instead of a Bell state we require tripartite entangled state as quantum channel. As mentioned in Section \ref{intro}, many of the papers which involve multiqubit complex states for teleportation also perform controlled teleportation using those multiqubit complex states. Here, we extend our scheme, for optimal QT to optimal CT. The scheme of CT using optimal resources can be explained along the same line of the QT scheme as follows. 

To construct an optimal scheme, it is assumed that Charlie also knows the unitary Alice is using to reduce the size of the quantum state to be teleported. In other words, he is aware of the number of entangled qubits Alice and Bob require to perform teleportation. Suppose the reduced quantum state has $m^\prime$ qubits, Charlie prepares $m^\prime$ GHZ states and share the three qubits among Alice, Bob and himself. Charlie measures his qubit in $\{\ket{+},\ket{-}\}$ basis and withholds the measurement outcome. Independently, Alice and Bob perform the QT scheme with the only difference that Bob requires Charlie's measurement disclosure to reconstruct the state. Charlie announces the required classical information when he wishes Bob to reconstruct the state. 

Similarly, when both Alice and Bob wish to teleport a quantum 
state each to Bob and Alice, respectively, under the supervision of Charlie, they perform QT schemes independently, while Charlie prepares the quantum channel in such a way that after his measurement the reduced state is the product of $2m^\prime$ Bell states (half of which will be used for Alice to Bob, while the remaining half for Bob to Alice communication). Charlie's disclosure of his measurement outcomes end both Alice's and Bob's ignorance regarding the quantum channel they were sharing, and they can subsequently reconstruct the unknown quantum states teleported to them (see \cite{thapliyal2015general} for detail).
In Ref. \cite{thapliyal2015applications}, some of the present authors have shown that BCST can also be accomplished solely using Bell states. Therefore, CT and BCST can also be performed using only bipartite entanglement. In absence of the controller, a BCST scheme can be reduced to a BST scheme. 
In the light of our present results, some of the recent schemes of CT using 4-qubit cluster state \cite{song2008controlled} and QIS using 4 and 5-qubit cluster state \cite{muralidharan2008quantum,nie2011quantum}; BST  using 3-qubit GHZ state \cite{hassanpour2016bidirectional}; and 6-qubit cluster state \cite{li2016asymmetric}, can also be performed with reduced amount of quantum resources (entangled states involving lesser number of qubits).

\section {Experimental implementation of the proposed efficient QT scheme using  IBM's real quantum processor} \label{ex}
\begin{figure}
	\begin{center}		\includegraphics[scale=.95]{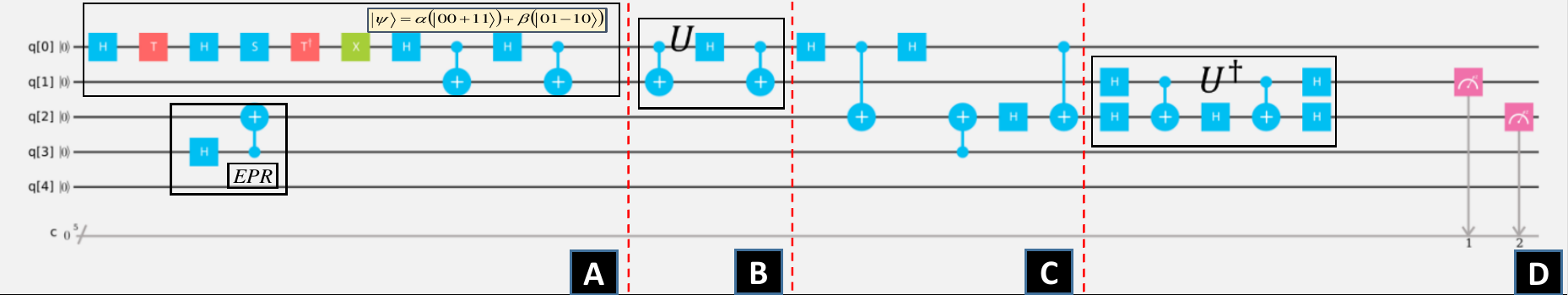}\end{center}
	\protect\caption{\label{fig:sim}(Color online) Teleportation circuit used on the IBM QE. Details of the various parts A-D are as follows. (A) Preparation of an unknown quantum state $\ket{\psi}$ with $\abs{\alpha}^2=0.375$ and $\abs{\beta}^2=0.125$ and the Bell state (EPR channel). (B) The unitary operation $U$ to transform 2-qubit state $\left(\alpha \left(\ket{00}+\ket{11}\right)+ \beta \left(\ket{01}-\ket{10}\right)\right)$ to state $\left(\alpha \ket{00}+ \beta \ket{10}\right)$ in the computational basis. (C) Teleportation of a single qubit state. (D) Reconstructing the state $\ket{\psi}$ from the teleported single qubit state by unitary operation $U^{\dagger}$ followed by projective measurement.}     
		\end{figure}

Recently, IBM corporation has placed a 5-qubit superconductivity-based quantum computer on cloud \cite{IBMQE}, and has provided its access to everyone. This initiative has enabled the interested researchers to experimentally realize various proposals for quantum information processing tasks. Interestingly, a set  of such implementations have already been reported \cite{fedortchenko2016quantum,rundle2016quantum,devitt2016performing}. Currently, this 5-qubit superconductivity-based quantum computer  has many limitations, like available gate library is only approximately universal, measurement of individual qubits at different time points is not allowed, limited applicability of C-NOT gate, and short decoherence time \cite{IBMDT}.  Further, the real quantum computer  (IBM quantum experience) available at cloud allows a user to perform an experiment using  at most 5-qubits. Keeping these limitations in mind, we have chosen a 2-qubit two-unknown quantum state $\ket{\psi}=\alpha(\ket{00}+\ket{11})+\beta(\ket{01}-\ket{10})$ : $2(|\alpha|^2+|\beta|^2)=1$ as the state to be teleported. Here, we would like to mention that implementation of a single qubit teleportation protocol (which requires only three qubits) has already been demonstrated using IBM's quantum computer \cite{fedortchenko2016quantum}. The experimental implementation of the present QT scheme is relatively complex and can be divided into four parts as shown in Figure \ref{fig:sim}. Part A involves preparation of state $\ket{\psi}$ (using qubit q[0] and q[1]) and a Bell state (using qubit q[2] and q[3]). The complex circuit comprised of the quantum gates from Clifford group is shown in Figure \ref{fig:sim}. Here, it may be noted that the IBM quantum computer accepts quantum gates from Clifford group only.  The state $\ket{\psi}$ in this particular case is prepared with $|\alpha|^2=0.375$ and $|\beta|^2=0.125$. Preparation of the desired 2-qubit state by application of specific quantum gates is detailed in the following.

$\ket{00}
\xrightarrow{\hspace{0.2cm}{\mathrm{H}^{1},\,\mathrm{T}^{1}}\hspace{0.2cm}}  \left(\frac{\ket{0}+ e^{i\frac{\pi}{4}} \ket{1}}{\sqrt{2}}\right) \ket{0}\xrightarrow{\hspace{0.2cm}{\mathrm{H}^{1},\,\mathrm{S}^{1}}\hspace{0.2cm}} e^{i\frac{\pi}{8}}(\cos\left({\frac{\pi}{8}}\right)\ket{0}+\sin\left({\frac{\pi}{8}}\right)\ket{1}) \ket{0} 
\xrightarrow{\hspace{0.2cm}{{\mathrm{T}^{\dagger}}^{1},\,\mathrm{X}^{1},\,\mathrm{H}^{1}}\hspace{0.2cm}} \sqrt{2}e^{i\frac{\pi}{8}}({\alpha \ket{0}+\beta \ket{1}}) \ket{0} \xrightarrow{\hspace{0.2cm}{\mathrm{C^{1}-NOT^{2}}}\hspace{0.2cm}}
\sqrt{2}e^{i\frac{\pi}{8}}(\alpha \ket{00}+\beta \ket{11}) \xrightarrow{\hspace{0.2cm} \mathrm{H}^{1}\hspace{0.2cm}}  e^{i\frac{\pi}{8}}\left(\alpha \left(\ket{00}+\ket{10}\right)+\beta \left(\ket{01}-\ket{11}\right)\right)
\xrightarrow{\hspace{0.2cm}\mathrm{C^{1}-NOT^{2}} \hspace{0.2cm}}e^{i\frac{\pi}{8}}\left(\alpha\left(\ket{00}+\ket{11}\right)+\beta\left(\ket{01}-\ket{10}\right)\right)$. 

Here, $e^{i\frac{\pi}{8}}$ is the global phase in the simulated quantum state with $\alpha= \frac{1}{\sqrt{2}}\left(\cos{\frac{\pi}{8}}+e^{-i\frac{\pi}{4}}\sin{\frac{\pi}{8}}\right) $, and
$\beta=\frac{1}{\sqrt{2}}\left(-\cos{\frac{\pi}{8}}+e^{-i\frac{\pi}{4}}\sin{\frac{\pi}{8}}\right)$. We have explicitly mentioned the qubit-number on which a particular operation is to be performed by mentioning the qubit-number on the the superscript of the corresponding unitary operator.

As described in Sec. \ref{2un}, Part B involves application of a unitary $U= \left(\mathrm{C^{2}-NOT^{1}}\right) \cdot \left(\mathbb{I}\otimes H\right) \cdot \left(\mathrm{C^{2}-NOT^{1}}\right)$ to transform the state $\ket{\psi}$ from the entangled basis to the computational basis and is the bottleneck of the protocol. Such a transformation allows us to render the information encoded into a smaller number of qubits (in our example it is a single qubit) and thus it reduces the amount of resources required. Part C is dedicated to the teleportation of a single qubit state. Here, we have used computational counterpart of teleportation \cite{adami1999quantum}, which can be performed when both Alice's and Bob's qubits are locally available for a 2-qubit operation. Teleportation of a single qubit state in analogy of Ref. \cite{fedortchenko2016quantum} can also be performed. This part of the circuit can be divided into two sub-parts. The first one (left aligned), which includes an EPR circuit, entangles qubit q[1] to the Bell state while the second part (right aligned) disentangles Bob's qubit (q[3]) from Alice's qubits. The need of disentangling Bob's qubit from Alice's qubit is explained above. In the standard protocol for QT  \cite{bennett1993teleporting}, Alice measures her qubits and announces measurement outcomes. Depending on the measurement outcome of Alice, Bob applies a unitary operation and reconstructs the unknown state. In IBM's quantum computer, simultaneous measurement of all the qubits is mandatory, which will project Bob's qubit into a mixed state. Therefore, we preferred to disentangle Bob's qubit from Alice's qubits before measurement. At last, in Part D, Bob applies the unitary $U^{\dagger}$ followed by the projective measurement on all qubits, which reveals the state $\ket{\psi}$ teleported to Bob's qubits.

	\begin{figure}
		\begin{center}		\includegraphics[scale=0.65]{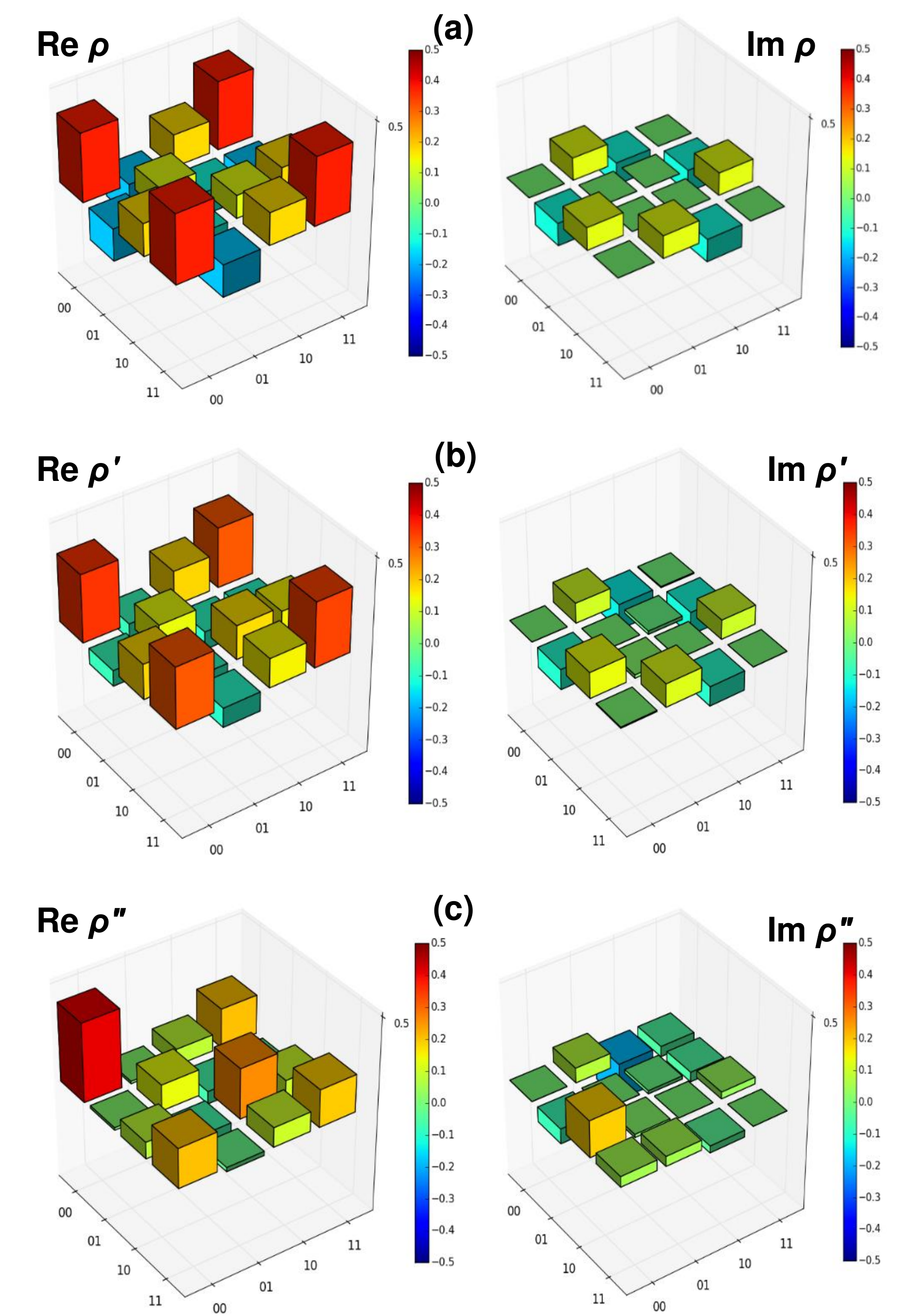}\end{center}\protect\caption{\label{fig:tom}(Color online) Graphical representation of real (Re)  and imaginary (Im)  parts of the density matrices of (a) the theoretical state $\alpha(\ket{00}+\ket{11})+\beta(\ket{01}-\ket{10})$, (b) the experimentally prepared state, and (c) the reconstructed state after teleportation.}
	\end{figure}

To perform a quantitative analysis of the performance of the QT scheme under consideration, we would require the density matrices of the state to be teleported and that of the teleported state. In a recent implementation of QT on IBM computer only probabilities of various outcomes were obtained \cite{fedortchenko2016quantum}. However, to obtain the full picture, we need to reconstruct the density matrix of the teleported state using quantum state tomography \cite{chuang1998bulk}. Till date, a large number of advanced protocols have been proposed for quantum state tomography \cite{schmied2016quantum}. An advanced protocol suppress the requirement of repeated preparation of the state to be tomographed and hence allows state characterization in dynamical environment using only one experiment \cite{shukla2013ancilla}. Here, we use  the method proposed by Chuang et al. \cite{chuang1998bulk}. According to which  we would require to obtain fifteen unknown parameters in a 2-qubit density matrix, and the same can be obtained using nine measurements. Using quantum state tomography (see \cite{james2001measurement,hebenstreit2017compressed,alsina2016experimental,rundle2016quantum,filipp2009two} for detail), we have reconstructed the teleported state (using nine rounds of experiments with 8192 runs of each experiment) as 
 \begin{equation}
 \rho^{\prime\prime}=\left[{\begin{array}{cccc}
	0.41 & 0.0125 + 0.0775 i & 0.085 - 0.19 i & 0.2035 - 0.05425 i\\
	0.0125 - 0.0775 i & 0.134 & -0.0645 - 0.01625i & -0.021 - 0.051 i\\
	0.085+ 0.19 i & -0.0645 + 0.01625 i & 0.261 & 0.101 + 0.0355 i\\
	0.2035 + 0.05425 i & -0.021 + 0.051 i  & 0.101 - 0.0355 i & 0.195\\
	\end{array}}\right],
\end{equation}
whereas theoretically the state prepared for the teleportation is $\rho=\ket{\Psi}\bra{\Psi}$ with $\ket{\Psi}=
\left\{\alpha \left(\ket{00}+\ket{11}\right)+ \beta \left(\ket{01}-\ket{10}\right)\right\}.$

During experimental implementation the state prepared may also have some errors. Keeping this in mind, we have reconstructed the density matrix of the quantum state (which is to be teleported) generated in the experiment  as
             
  \begin{equation}
   \rho^\prime=\left[{\begin{array}{cccc}
	0.352 & -0.0805 + 0.1045 i & 0.18- 0.133 i & 0.31325- 0.005 i\\
	-0.0805 - 0.1045 i & 0.135 & -0.09275 - 0.017 i & -0.099 - 0.12 i\\
	0.18+ 0.133 i & -0.09275 + 0.017 i & 0.175 & 0.1505 + 0.1005 i\\
	0.31325 + 0.005 i & -0.099 + 0.12 i  & 0.1505 - 0.1005 i & 0.338\\
	\end{array}}\right].
\end{equation}

Various elements of all the density matrices are shown pictorially in Figure \ref{fig:tom}.   
  Finally, we would like to quantize the performance of the QT scheme using a distance based measure, fidelity, defined as  $F=Tr[\sqrt{\sqrt{\rho_1}.\rho_2.\sqrt{\rho_1}}].$ Using this we calculated the fidelity between the theoretical state with experimentally generated state (i.e., $\rho_1=\rho$ and $\rho_2=\rho^\prime$) as 0.9221. The same calculation performed between experimentally generated and teleported state (i.e., $\rho_1=\rho$ and $\rho_2=\rho^{\prime\prime}$) yields a higher value for fidelity (0.9378). Thus, the state preparation is relatively  more erroneous, due to errors in gate implementation and decoherence. However, the constructed state is found to be teleported with high fidelity.

\section {Conclusion} \label{con}
Teleportation of multi-qubit states with the optimal amount of quantum resources  in terms of the number of entangled qubits required in the quantum channel has been performed.  Specifically, it has been shown that the amount of quantum resources required to teleport an unknown quantum state depends only on the number of non-zero probability amplitudes in the quantum state and is independent of the number of qubits in the state to be teleported.

The present study establishes a  foundationally important fact that the more the information regarding the quantum state to be teleported is available the lesser the quantum resources to teleport that state should be required. Thus, there exists a trade-off between our knowledge about the state to be teleported and the amount of quantum resources required for the teleportation.  Every scheme of RSP that performs teleportation of a known state using lesser amount of quantum resources than that required for teleportation of an unknown quantum state \cite{pati2000minimum,sharma2015controlled}, essentially exploits this trade-off.

Further, the relevance of the present work is not restricted to QT. It is also useful in CT, BST and BCST schemes. The relevance of the present work also  lies in the fact that  the limiting cases of our scheme can perform the same task with reduced amount of quantum resources in comparison  with the previously achieved counterparts. In fact, for almost all the existing works reported on teleportation of multi-qubit states with some non-zero unknowns, we have shown a clear prescription to optimize the required quantum resources.

Finally, a proof of principle experimental implementation of the proposed scheme on the  IBM quantum computer is shown for teleporation of a 2-qubit quantum state. It is important to note that our scheme enables this experimental realization to be accomplished on a 5-qubit quantum computer. From this implementation,  we have computed  the fidelity between the theoretical and experimentally generated state, and experimentally generated and teleported state. This quantitative analysis infer that the teleportion circuit implemented here is more efficient when compared with the state preparation part. This fact establishes the relevance of the proposed scheme in context of reduction of  the decoherence effects on teleportation, too.

We hope our attempt to optimize the resource requirement for teleportation of multi-qubit quantum states should increase the feasibility of multi-qubit quantum state teleportation performed in various quantum systems. This is also expected to impact the teleportation-based direct secure quantum communication scheme, where resources can be optimized exploiting the form of the quantum state teleported (e.g., \cite{joy2017efficient} and references therein). Along the same line, optimization of quantum resources in CT, without affecting the controller's power, will be performed  and reported elsewhere.

\textbf{Acknowledgment:}
AP, AS and KT thank Defense Research \& Development Organization (DRDO), India for the support provided through the project number ERIP/ER/1403163/M/01/1603. 

\bibliographystyle{Final}
\bibliography{ref1}

\end{document}